%% file: main.tex
\documentclass[journal,onecolumn,twoside]{IEEEtran}
\input{preamble}
\begin{document}

\title{Almost-Isometric Embeddings of Graphs into Riemannian Manifolds}

\input{author_info}
\maketitle

\input{sections/abstract}
\input{sections/introduction}
\input{sections/related_work}
\input{sections/preliminaries}
\input{sections/main_result}
\input{sections/proof_overview}
\input{sections/spatial_embeddings}
\input{sections/surface_realization}
\input{sections/space_information_flow}
\input{sections/conclusion}
\input{sections/acknowledgment}

\input{references}
\end{document}

%% file: preamble.tex
\usepackage{amsmath,amsfonts}
\usepackage{amsthm}
\usepackage{algorithmic}
\usepackage{algorithm}
\usepackage{array}
\usepackage[caption=false,font=normalsize,labelfont=sf,textfont=sf]{subfig}
\usepackage{textcomp}
\usepackage{stfloats}
\usepackage{url}
\usepackage{verbatim}
\usepackage{graphicx}
\usepackage{float}
\usepackage{cite}
\usepackage[utf8]{inputenc} 
\usepackage[T1]{fontenc}
\usepackage{ifthen}
\usepackage{amssymb}
\DeclareMathOperator{\Route}{Route}
\DeclareMathOperator{\Code}{Code}
\DeclareMathOperator{\Adv}{Adv}

\theoremstyle{plain}
\newtheorem{theorem}{Theorem}[section]
\newtheorem{lemma}[theorem]{Lemma}

\newtheorem{corollary}[theorem]{Corollary}

\theoremstyle{definition}

\theoremstyle{remark}

\input{figures/style}

\input{figures/captions}
\input{figures/placement}

%% file: figures/style.tex
\usepackage{tikz}
\usetikzlibrary{arrows.meta,calc}
\providecommand{\MFChinese}{0}
\newcommand{\mftext}[2]{\ifnum\MFChinese=1\relax#2\else#1\fi}
\definecolor{mfInk}{HTML}{242838}
\definecolor{mfEdge}{HTML}{353F53}
\definecolor{mfBlue}{HTML}{64748F}
\definecolor{mfLight}{HTML}{CFD6E2}
\definecolor{mfPale}{HTML}{F3F5FA}
\definecolor{mfRust}{HTML}{A3442D}
\definecolor{mfIndigo}{HTML}{3E438A}
\definecolor{mfGray}{HTML}{586173}
\definecolor{mfRule}{HTML}{C3C8D2}
\tikzset{
  mf diagram/.style={x=1cm,y=1cm,line cap=round,line join=round,
    every node/.style={inner sep=1.4pt,text=mfInk,font=\normalfont\footnotesize}},
  mf heading/.style={font=\normalfont\small,anchor=west},
  mf detail/.style={font=\normalfont\scriptsize},
  mf edge/.style={draw=mfEdge,line width=1.05pt},
  mf axis/.style={draw=mfGray,line width=.55pt,dash pattern=on 2pt off 1.5pt},
  mf lead/.style={draw=mfGray,line width=.55pt},
  mf guide/.style={draw=mfRule,line width=.6pt},
  mf step/.style={-{Stealth[length=2.3mm,width=1.55mm]},draw=mfGray,line width=1pt},
  mf forward/.style={-{Stealth[length=2.1mm,width=1.45mm]},draw=mfIndigo,line width=1pt},
  mf back/.style={-{Stealth[length=2.1mm,width=1.45mm]},draw=mfRust,line width=1pt},
  mf vertex/.style={circle,draw=mfEdge,fill=white,line width=.9pt,inner sep=0pt,minimum size=4.8pt},
  mf mapped/.style={circle,draw=white,fill=mfIndigo,line width=.65pt,inner sep=0pt,minimum size=5.3pt},
  mf relay/.style={circle,draw=white,fill=mfRust,line width=.7pt,inner sep=0pt,minimum size=6.3pt},
  mf ring/.style={draw=mfBlue,line width=.65pt},
  mf warm/.style={text=mfRust},mf cool/.style={text=mfIndigo}
}


%% file: figures/captions.tex
\newcommand{\mfConstructionCaption}{\mftext{%
Construction of the surface \(M_r\). Each edge of the crossing-free drawing \(\Gamma_0\) is lengthened in its own cylinder. Only the deformation of edge \(e=\{1,3\}\) is shown. Thickening the resulting curve gives a tube with the same wavy axis. The example has \(n=4\), \(m=5\), and genus \(2\).}{%
曲面 \(M_r\) 的三步构造。无交叉直线嵌入 \(\Gamma_0\) 的各条边分别在互不相交的圆柱内调长。图中仅示出边 \(e=\{1,3\}\) 的形变；加粗后，圆管沿同一波浪形轴线延伸。例中 \(n=4\)、\(m=5\)，所得曲面亏格为 \(2\)。}}
\newcommand{\mfTubeCaption}{\mftext{%
Local geometry and the two maps. (a) Trimming near \(p_u,p_v\) leaves a core of arclength \(\ell_e=L_e-O(r)\). The dashed line is the core \(\gamma_e\). Its translate \(X_e\) in the direction normal to the edge plane has the same length, even though the inside of a bend is shorter. (b) \(\iota\) traverses \(\lambda_e\) at constant speed, including the connecting arcs in both junctions. Under \(\pi\), each entire junction maps to its vertex and a section at \(s\) maps to the point at distance \(c_e s/\ell_e\) from \(u\) along \(e\).}{%
薄管的局部几何与双向映射。(a) 在 \(p_u,p_v\) 附近截短后，中心弧长为 \(\ell_e=L_e-O(r)\)。虚线表示中心弧 \(\gamma_e\)。沿空间边所在平面的法向平移得到的管壁曲线 \(X_e\) 与中心弧等长，而弯管内侧较短。(b) \(\iota\) 沿 \(\lambda_e\) 匀速行进，包括两端接合面内的连接弧。\(\pi\) 将每个接合面整体压回相应顶点，把轴向坐标为 \(s\) 的横截面压回边 \(e\) 上距 \(u\) 为 \(c_e s/\ell_e\) 的点。}}

%% file: figures/placement.tex
\makeatletter
\newcommand{\mfInsertFigure}[3]{%
 \if@twocolumn
   \begin{figure*}[!t]\centering
   \resizebox{.97\linewidth}{!}{\input{figures/#1}}\par
   \caption{#2}\label{#3}\end{figure*}%
 \else
   \begin{figure}[!htbp]\centering
   \resizebox{.97\linewidth}{!}{\input{figures/#1}}\par
   \caption{#2}\label{#3}\end{figure}%
 \fi}
\makeatother

%% file: author_info.tex
\author{Sirui~Liu,~\IEEEmembership{Student Member,~IEEE,}
        Zongpeng~Li,~\IEEEmembership{Senior Member,~IEEE,}
        and~Baochun~Li,~\IEEEmembership{Fellow,~IEEE}
\thanks{Sirui Liu is with Institute for Network Sciences \& Cyberspace, Tsinghua University, Beijing, China, and also with Quan Cheng Laboratory, Shandong, China. Email: liu-sr24@mails.tsinghua.edu.cn.}
\thanks{Zongpeng Li is with Quan Cheng Laboratory, Shandong, China, and also with Institute for Network Sciences \& Cyberspace, Tsinghua University, Beijing, China.  Email: zongpeng@tsinghua.edu.cn. Corresponding author.}
\thanks{Baochun Li is with University of Toronto, Canada. Email: bli@ece.toronto.edu.}
}

%% file: sections/abstract.tex
\begin{abstract}
Graphs represent objects as vertices and relationships as edges; positive edge weights define shortest-path metrics. Metric embeddings leverage target-space geometry to analyze distances. Foundational results target linear spaces: Nash embedded Riemannian manifolds isometrically into Euclidean space, and Bourgain embedded $n$-point metrics into Hilbert space with $O(\log n)$ distortion. By contrast, the manifold hypothesis places high-dimensional data near low-dimensional smooth manifolds, making such manifolds themselves target spaces. This naturally asks: with what distortion can finite graph metrics embed into low-dimensional smooth manifolds under intrinsic distance? Although the star $K_{1,3}$ cannot embed isometrically into any complete Riemannian manifold, we prove almost-isometric embeddings always exist: for $\varepsilon>0$ and any finite connected positively weighted graph, we construct a smooth connected closed orientable surface in $\mathbb{R}^3$ admitting a noncontractive $(1+\varepsilon)$-embedding of its vertices, with $\varepsilon$ arbitrarily small and independent of graph parameters. This construction transfers communication schemes between graphs and surfaces. For independent unicasts, fractional routing splits flows across paths while network coding lets relays combine messages. Space information flow permits relays anywhere, charging bandwidth times distance. Whereas coding advantage (optimal routing over coding cost) equals one in Euclidean space, we prove coding advantages on finite positively weighted undirected graphs transfer to smooth closed surfaces within factors approaching one. Counterexamples to the multiple-unicast conjecture and amplification yield smooth closed surfaces with unbounded coding advantage. Conversely, coding advantage is at most $C\log(g+2)$ on finite undirected graphs of orientable genus $g$ and smooth closed orientable surfaces of genus $g$, for absolute $C$.
\end{abstract}


%% file: sections/introduction.tex
\section{Introduction}\label{sec:introduction}

Graphs represent objects by vertices and relationships by edges~\cite{BollobasModernGraphTheory,DiestelGraphTheory}. In a connected graph with positive edge weights, the distance between two vertices is the length of a shortest path joining them. Such shortest-path metrics arise throughout combinatorics, theoretical computer science, information theory, and network science, and have a direct meaning in communication networks: when edge weights represent link transmission costs, shortest paths determine minimum-cost routes. One approach to analyzing discrete distances is to map a metric space into a target space with geometric or algebraic structure while controlling distortion, which measures the relative change in pairwise distances, so that analytic and geometric tools of the target space apply to the original space.

Many foundational embedding theorems across both continuous and discrete geometry target linear spaces.
On the continuous side, Riemann developed the intrinsic geometry of manifolds from within: a Riemannian metric defines the arc length of curves on a manifold, and the infimum of curve lengths between two points gives their intrinsic distance.
Nash~\cite{Nash1954,Nash1956} subsequently proved that every smooth Riemannian manifold embeds isometrically into sufficiently high-dimensional Euclidean space.
On the discrete side, Bourgain~\cite{Bourgain1985} proved that every $n$-point metric space embeds into Hilbert space with distortion $O(\log n)$, with subsequent work establishing distortion bounds for embedding graph metrics into various normed spaces~\cite{LinialLondonRabinovich1995,DezaLaurent1997,Matousek2002}.
Whereas these results target linear spaces, the manifold hypothesis in machine learning and high-dimensional statistics posits that high-dimensional data concentrate near low-dimensional smooth manifolds~\cite{FeffermanMitterNarayanan2016,BengioCourvilleVincent2013}.
Under this hypothesis, the analysis of high-dimensional data can focus on the underlying manifold rather than being governed by the ambient dimension.
To determine what constraints the manifold hypothesis imposes on finite samples, one must identify which finite metric spaces embed into low-dimensional smooth Riemannian manifolds equipped with their intrinsic distance; since every finite metric is the shortest-path metric of a finite positively weighted graph, this is equivalent to characterizing which finite positively weighted graph metrics admit such embeddings---a question that has not yet been systematically answered.

If an isometric embedding without distortion is required, smoothness of the target metric rules out many finite graphs. Although the Kuratowski map embeds every finite metric space isometrically into $\ell_\infty$~\cite{DezaLaurent1997,Kuratowski1935}, its $\ell_\infty$ norm is nonsmooth. On a complete connected smooth Riemannian manifold, any two points can be joined by a curve whose length equals their intrinsic distance. Such a curve, called a minimizing geodesic when parametrized by arc length, continues smoothly in its tangent direction and cannot turn a corner or branch. This directly rules out an isometric embedding of the four-vertex star $K_{1,3}$: its three leaves are pairwise at distance $2$ via the center, so in any isometric embedding concatenating the unit-length minimizing geodesics through the center for any pair of leaves would give a length-$2$ minimizing geodesic. The three geodesics from the center to the leaves would therefore point in pairwise opposite directions there, which is impossible for three unit vectors. Since isometric embeddings fail already for $K_{1,3}$, a natural question arises: once multiplicative distortion greater than $1$ is allowed, does the shortest-path metric of every finite connected positively weighted graph admit an almost-isometric embedding into a smooth low-dimensional Riemannian manifold equipped with its intrinsic distance?

We answer this question in the affirmative. Let $G=(V,E,c)$ be a finite connected positively weighted graph with $n$ vertices and $m$ edges. For every $\varepsilon>0$, we construct a smooth closed connected orientable surface $M\subset\mathbb{R}^3$ of genus $m-n+1$ and an injective map $f:V\to M$ such that the intrinsic distance on $M$ satisfies

\begin{equation}
  d_G(u,v)\le d_M\bigl(f(u),f(v)\bigr)\le(1+\varepsilon)d_G(u,v), \qquad u,v\in V.\label{eq:vertex_intro}
\end{equation}

Intuitively, such a surface is closed without boundary, and its genus counts the number of handles (for instance, genus $0$ for a sphere and genus $1$ for a torus; see Section~\ref{subsec:smooth_surfaces} for formal definitions). Since every finite metric is the shortest-path metric of the complete graph weighted by its pairwise distances, \eqref{eq:vertex_intro} shows that every finite metric space embeds with distortion arbitrarily close to $1$ into a two-dimensional smooth closed surface whose topology and Riemannian metric may depend on the data. For the manifold hypothesis to constrain finite intrinsic distance data, the candidate manifolds must therefore satisfy further conditions. For example, the statistical model of Fefferman {\em et al.}~\cite{FeffermanMitterNarayanan2016} imposes an upper bound on manifold volume and a lower bound on reach.

This construction also enables transferring communication schemes between graphs and smooth surfaces: in communication networks, positive edge weights represent link costs per unit bandwidth. When multiple independent source-sink pairs communicate simultaneously over a finite undirected network, fractional routing splits each session's flow across paths, whereas network coding also allows intermediate nodes to combine incoming messages algebraically. The multiple-unicast conjecture of Li and Li~\cite{LiLi2004} asserted that network coding cannot augment the achievable rate region of fractional routing under identical edge capacities---an assumption used in conditional lower bounds for external-memory integer sorting and constant-degree multiplication circuits~\cite{FarhadiEtAl2019,AfshaniEtAl2019}---until recent counterexamples by Zhang {\em et al.}~\cite{ZhangLiLi2026} and by Braverman and He~\cite{BravermanHe2026} disproved the conjecture on finite undirected graphs. Extending this comparison to continuous geometry rests on an equivalence between throughput and cost: the multiple-unicast conjecture holds if and only if network coding has no cost advantage over fractional routing on undirected multiple-unicast networks, where total network cost sums edge flows weighted by edge lengths.

Li and Wu~\cite{LiWu2012SIF,LiWu2012SIFReport} introduced the space information flow model to formulate this minimum-cost network layout problem in a continuous metric space, and Xiahou {\em et al.}~\cite{XiahouLiWuHuang2014} subsequently developed geometric network design in this setting: terminals are fixed in the space, any finite number of relay and coding nodes may be placed at arbitrary points to lay out links whose cost equals bandwidth times distance, and the objective is to minimize total link cost. For independent unicast sessions in Euclidean space~\cite{LiWu2012SIF} and in $\ell_p^n$ for $1\le p<2$~\cite{LiuLi2025}, optimal network coding cost and optimal fractional routing cost are always equal. Xiahou {\em et al.}~\cite{XiahouLiWuHuang2014} raised the question of multiple-unicast coding advantage on Riemannian manifolds, but for more than a decade after space information flow was introduced, it remained open whether optimal network coding cost could be lower than optimal fractional routing cost on a smooth manifold.

Establishing a correspondence between space information flow on discrete graphs and on smooth closed surfaces requires more than controlling vertex distances, because relays may be placed anywhere on the surface and links can be arbitrarily short. Although Gromov--Hausdorff approximations by smooth surfaces in $\mathbb{R}^3$~\cite{Cassorla1992} yield almost-isometric embeddings of the vertex set, they neither control the surface genus nor avoid additive distance errors, which cannot produce a uniform multiplicative cost bound when arbitrarily short relay links accumulate. Instead, viewing $G$ as the metric graph $|G|$ obtained by gluing intervals of lengths $c_e$, we embed $G$ without crossings into $\mathbb{R}^3$, take the boundary of a thin tubular neighborhood as a smooth closed surface $M$ of genus $m-n+1$, and directly construct continuous maps $\iota:|G|\to M$ and $\pi:M\to|G|$ satisfying

\[
  \operatorname{Lip}(\iota)\le1+\varepsilon,\qquad \operatorname{Lip}(\pi)\le1,\qquad \pi(\iota(v))=v,\quad v\in V.
\]

Here $\iota$ increases the cost of any graph solution by at most a factor of $1+\varepsilon$, while the globally nonexpansive map $\pi$ projects any surface layout with arbitrarily placed relays and coding nodes back to $|G|$ link by link without increasing cost. Expanding these links along shortest paths and contracting each unbranched chain of segments back to its original edge gives a scheme on the discrete graph $G$ of no greater cost. Consequently, for all finite multiple-unicast demand families on $V$ simultaneously, both optimal fractional routing cost and optimal network coding cost on $M$ lie between $1$ and $1+\varepsilon$ times their graph counterparts.

Thus, for every nonempty demand family on the graph vertices, the ratio of the coding advantages before and after transfer to $M$ can be made arbitrarily close to $1$. Combining recent counterexamples to the multiple-unicast conjecture~\cite{ZhangLiLi2026,BravermanHe2026} with the amplification of the common-rate gap~\cite{BravermanGargSchvartzman2017} and throughput--cost duality~\cite[Theorem~4.1]{XiahouLiWuHuang2014} gives finite positively weighted undirected graph instances with arbitrarily large coding advantage, and hence a family of smooth closed surfaces with unbounded coding advantage. In this family, both the graph size and the number of independent cycles tend to infinity, and the genus of each constructed surface equals the number $m-n+1$ of independent cycles. This raises the complementary question of whether achieving arbitrarily large coding advantage on smooth closed surfaces necessarily requires high genus. We answer this question affirmatively. For graphs, we use the orientable genus, the least genus of a closed orientable surface on which the graph can be drawn without crossings. We prove that coding advantage is at most $C\log(g+2)$ on both finite positively weighted undirected graphs of orientable genus at most $g$ and smooth connected closed orientable surfaces of genus at most $g$, where $C$ is an absolute constant. Thus, achieving coding advantage $A$ requires genus $g\ge\exp(A/C)-2$.

Although minimizing geodesics on a smooth manifold cannot turn a corner or branch at an interior point, the maps $(\iota,\pi)$ allow us to preserve graph vertex distances on a two-dimensional smooth closed surface with distortion arbitrarily close to $1$ while controlling the lengths of paths transferred in either direction between the entire metric graph $|G|$ and the surface. The space information flow result illustrates the broader significance of this construction. More broadly, for other problems defined in terms of distances or path lengths, the same length bounds may help relate the corresponding quantities on the graph and the surface whenever feasible structures transfer in both directions under $(\iota,\pi)$.

Our main contributions are threefold.

\begin{enumerate}
\item \textbf{Almost-isometric embeddings of finite metric spaces into smooth closed surfaces.} For every finite connected undirected graph $G$ with $n$ vertices and $m$ positively weighted edges and every $\varepsilon>0$, we construct a smooth closed connected orientable surface $M$ of genus $m-n+1$ in $\mathbb{R}^3$, a $(1+\varepsilon)$-Lipschitz map $\iota:|G|\to M$ from the metric graph to the surface, and a nonexpansive map $\pi:M\to|G|$ such that $\pi\circ\iota$ fixes every vertex. The intrinsic distance between the images of any two vertices lies between their graph distance and $1+\varepsilon$ times that distance. The complete-graph representation gives such an almost-isometric embedding for every finite metric space.

\item \textbf{Transfer of coding advantage in space information flow.} In multiple-unicast communication where source-sink pairs transmit independent messages, fractional routing splits and forwards flows along paths whereas network coding allows intermediate nodes to combine messages algebraically; measuring total cost by the sum of link bandwidths times distances, coding advantage is the ratio of optimal fractional routing cost to optimal network coding cost. We prove that coding advantage transfers from finite positively weighted undirected graphs to smooth closed surfaces in $\mathbb{R}^3$, with the ratio of the two coding advantages arbitrarily close to $1$. Applying this transfer to known undirected graph instances with arbitrarily large coding advantage yields a family of smooth closed surfaces with unbounded coding advantage.

\item \textbf{Genus upper bound on coding advantage for graphs and surfaces.} The orientable genus of a graph is the minimum genus of a closed orientable surface into which the graph embeds without crossings. For any finite connected positively weighted undirected graph of orientable genus at most $g$, or any smooth closed connected orientable Riemannian surface of genus at most $g$ with unrestricted relay placement, the optimal fractional routing cost for any finite family of independent unicast demands is at most $C\log(g+2)$ times the optimal network coding cost, where $C$ is an absolute constant.
\end{enumerate}

Section~\ref{sec:related_work} reviews related work, and Section~\ref{sec:preliminaries} gives the preliminaries. Section~\ref{sec:main_result} states the main results, and Section~\ref{sec:proof_overview} outlines their proofs. Section~\ref{sec:spatial_embeddings} constructs spatial graph embeddings with prescribed edge lengths, and Section~\ref{sec:surface_realization} constructs the smooth surface and the maps in both directions. Section~\ref{sec:space_information_flow} studies coding advantage in space information flow, and Section~\ref{sec:conclusion} concludes the paper.

%% file: sections/related_work.tex
\section{Related Work}\label{sec:related_work}

\subsection{Finite Metric Embeddings}

When a finite metric space is mapped into a normed linear space, the geometry of the target space governs the achievable distortion. Without smoothness requirements, the Kuratowski map embeds every finite metric space isometrically into finite-dimensional $\ell_\infty$~\cite{DezaLaurent1997,Kuratowski1935}; for Hilbert space, Schoenberg's criterion states that an isometric embedding exists if and only if the squared-distance matrix is conditionally negative semidefinite~\cite{Schoenberg1935}. Allowing multiplicative distortion, Johnson and Lindenstrauss~\cite{JohnsonLindenstrauss1984} established dimension reduction for finite subsets of Hilbert space, and Bourgain~\cite{Bourgain1985} proved that every $n$-point metric space embeds into Hilbert space with $O(\log n)$ distortion; for comprehensive surveys, see~\cite{DezaLaurent1997,Matousek2002}.

Following this line, Linial {\em et al.}~\cite{LinialLondonRabinovich1995} studied embeddings of finite metrics into $L_1$. The connection to network flow comes from cut metrics, which give distance $0$ to vertices on the same side of a cut and $1$ to vertices on opposite sides. Because every finite $L_1$ metric is a nonnegative linear combination of these cut metrics~\cite{DezaLaurent1997}, a uniform bound on $L_1$ distortion over all positive edge lengths on an undirected graph bounds the ratio of the sparsest cut value to the maximum concurrent flow value on that graph~\cite{GuptaEtAl2004}. Every $n$-point metric embeds into $L_1$ with $O(\log n)$ distortion~\cite{Bourgain1985,LinialLondonRabinovich1995}, whereas on bounded-genus graphs the bound does not grow with the number of vertices: OpenAI~\cite{OpenAIPlanar2026} proved that shortest-path metrics of finite connected positively weighted planar graphs embed into $L_1$ with constant distortion, which combined with Sidiropoulos's stochastic planarization theorem~\cite{Sidiropoulos2010} gives $O(\log(g+2))$ distortion into $L_1$ for finite connected positively weighted graphs of orientable genus at most $g$. These embeddings into $\ell_\infty$, Hilbert space, and $L_1$ prescribe the type of linear target space and examine how distortion scales with vertex count or genus; by contrast, we consider embeddings into two-dimensional smooth closed surfaces whose genus and Riemannian metric may adapt to the input graph, achieving multiplicative distortion arbitrarily close to $1$.

\subsection{Metric Embeddings into Smooth Surfaces}

When a discrete set is embedded isometrically into a complete smooth Riemannian manifold, pairwise distances equal the lengths of minimizing geodesics. Because minimizing geodesics on a smooth manifold cannot bend sharply or branch, the finite star $K_{1,3}$ admits no isometric embedding into any complete smooth Riemannian manifold (see~\cite[proof of Theorem~2.1]{Ren2018}). For infinite discrete sets, Eilat and Klartag~\cite{EilatKlartag2021} proved that if a complete connected Riemannian surface contains an isometric copy of a planar net, a discrete set within a fixed distance of every point in the Euclidean plane, then the entire surface must be isometric to the Euclidean plane.

Allowing small distance errors leads to a different question: can a smooth surface approximate the entire metric graph, including the interiors of its edges? Distances measured along the edge intervals make a finite connected metric graph a compact inner metric space, so one can compare it with a surface in Gromov--Hausdorff distance~\cite{GromovMetricStructures,BuragoBuragoIvanov2001}. Cassorla~\cite{Cassorla1992} proved that in this distance every compact inner metric space can be approximated by smooth surfaces in $\mathbb{R}^3$. For a fixed finite graph, since distinct vertices have positive minimum separation, taking a sufficiently small additive error and rescaling the surface yields a noncontractive $(1+\varepsilon)$ embedding on the vertex set. However, this approximation does not control the genus of the surface and gives only an additive distance bound; when relays may be placed arbitrarily on the surface, summing per-link additive error bounds gives no uniform multiplicative bound on total network cost. To control the genus, one can instead start from a neighborhood of the graph in three dimensions. A thin tubular neighborhood of a crossing-free embedding of a finite connected graph with $n$ vertices and $m$ edges in $\mathbb{R}^3$ is a handlebody whose boundary is a smooth closed orientable surface of genus $m-n+1$~\cite{Hirsch1976,FriedlHerrmann2021}. These topological results establish the existence and topological type of such neighborhoods, without addressing how to prescribe arbitrary positive edge lengths in $\mathbb{R}^3$ or how to compare intrinsic surface distances quantitatively with shortest-path distances on the graph.

\subsection{The Manifold Hypothesis}

The manifold hypothesis, often positioned as the fundamental theory behind today's research on AI and large language models, posits that high-dimensional observations lie near a low-dimensional smooth manifold. Motivated by this view, nonlinear dimensionality-reduction methods extract low-dimensional coordinates from finite samples: Isomap~\cite{TenenbaumDeSilvaLangford2000}, for example, approximates geodesic distances on the manifold using shortest-path distances on a sample neighborhood graph, while locally linear embedding~\cite{RoweisSaul2000}, Laplacian eigenmaps~\cite{BelkinNiyogi2003}, and some representation learning methods~\cite{BengioCourvilleVincent2013} exploit local geometric structure in related ways.

When testing the manifold hypothesis or reconstructing a manifold quantitatively from finite samples, existing mathematical results supplement low-dimensional smoothness with bounds on volume, bending, and how closely distinct parts of a candidate manifold can approach one another. Fefferman {\em et al.}~\cite{FeffermanMitterNarayanan2016}, for instance, tested whether data lie near a manifold of bounded volume and reach bounded away from zero so that the manifold cannot bend too tightly or nearly intersect itself; results on recovering manifold topology from random samples~\cite{NiyogiSmaleWeinberger2008} and reconstructing Riemannian manifolds from finite or noisy intrinsic distances~\cite{FeffermanEtAl2020Whitney,FeffermanEtAl2020Noisy} likewise assume bounds on reach, curvature, or injectivity radius. Our embedding theorem shows that if such geometric bounds are omitted and the genus and metric of a two-dimensional smooth closed surface may depend on the data, every finite metric space embeds with distortion arbitrarily close to $1$; assuming only that data lie on a low-dimensional smooth manifold therefore does not restrict finite intrinsic distance structures up to arbitrarily small multiplicative distortion.

\subsection{The Multiple-Unicast Conjecture}

Network coding allows intermediate nodes to combine incoming messages algebraically before forwarding~\cite{AhlswedeCaiLiYeung2000,LiYeungCai2003,KoetterMedard2003}. For multiple independent unicast sessions on an undirected network, Li and Li~\cite{LiLi2004} and Harvey {\em et al.}~\cite{HarveyKleinbergLehman2004} posed the multiple-unicast conjecture, asserting that network coding and fractional routing share the same achievable rate region under identical edge capacities; by linear-programming duality, this is equivalent to network coding achieving no lower total transmission cost than optimal fractional routing for any given demands and nonnegative edge costs~\cite{LiLi2004}. Multicommodity flow results relate fractional routing feasibility to cut conditions for specific graph and terminal configurations~\cite{Hu1963,Seymour1980,OkamuraSeymour1981,Okamura1983,ChakrabartiEtAl2012}. For networks with few sessions, special terminal placements, or restricted topologies, coding upper bounds and graph reductions establish cases in which coding rates do not exceed fractional routing rates~\cite{JainVaziraniYuval2006,HarveyKleinbergLehman2006,YinEtAl2018,QureshiThakor2022,LiuEtAl2026,LiuLiFanChen2026Interaction,LiuQueLiLi2026BeyondCuts,LiuLiLi2026SingleSink}. In addition, Braverman {\em et al.}~\cite{BravermanGargSchvartzman2017} showed that whenever the common coding rate exceeds the common fractional routing rate on a single undirected graph, the ratio can be amplified across a family of graphs to grow polylogarithmically with the number of vertices.

The conjecture also served as an assumption for conditional lower bounds on external-memory integer sorting, constant-degree boolean multiplication circuits, and data structures~\cite{FarhadiEtAl2019,AfshaniEtAl2019,DvorakEtAl2021}. Several recent constructions refuted the conjecture: Zhang {\em et al.}~\cite{ZhangLiLi2026} used reversible circuits to build a binary linear coding counterexample on a subcubic unit-capacity undirected graph, and Braverman and He~\cite{BravermanHe2026} constructed an $\mathbb{F}_9$-linear counterexample on a bipartite graph. Li {\em et al.}~\cite{LiLiLiu2026Sorting} built on the finite network in an integer-multiplication construction~\cite{OpenAIMultiplication2026} and subsequent circuit refinements to construct a binary linear counterexample and improve upper bounds for external-memory integer sorting, while a grid arrangement of the XOR-swap unit from~\cite{OpenAIMultiplication2026} gave a 32-vertex counterexample~\cite{Dysfunctor2026}.

\subsection{Space Information Flow}

Space information flow places communication networks in a continuous metric space: source and sink terminals are fixed, relay and coding nodes may be placed anywhere in the space, and total network cost is defined as the sum of link bandwidths times distances~\cite{LiWu2012SIF,LiWu2012SIFReport}. For multiple independent unicast demands, Li and Wu~\cite{LiWu2012SIF} proved that optimal network coding cost equals optimal fractional routing cost in Euclidean space, and Liu and Li~\cite{LiuLi2025} extended this equality to $\ell_p^n$ for $1\le p\le 2$ via almost-isometric embeddings of finite subsets into $\ell_1$.

Xiahou {\em et al.}~\cite{XiahouLiWuHuang2014} used embeddings into finite-dimensional $\ell_p$ spaces to study multiple unicast on undirected graphs.
Their argument extends to any metric space with no coding advantage.
Suppose a map from the vertices of a graph into such a space preserves the shortest-path distance between each source and its sink.
If every edge has weight at least the distance between the images of its endpoints, then coding and fractional routing have the same optimal cost for those demands on the graph.
Xiahou {\em et al.} also asked whether network coding can cost less than fractional routing on a Riemannian manifold.

We answer this question by transferring a coding advantage in the reverse direction, from an undirected graph to a smooth closed surface.
By comparing costs in both directions between the metric graph and the entire surface, we show that the advantage persists even when relays may be placed anywhere on the surface.

%% file: sections/preliminaries.tex
\section{Preliminaries}\label{sec:preliminaries}

\subsection{Metric Graphs}

Let \(G=(V,E,c)\) be a finite connected undirected simple graph with positive edge lengths \(c:E\to\mathbb{R}_{>0}\), written \(c_e=c(e)\). Replacing each edge \(e\in E\) by a closed interval of length \(c_e\) and gluing its endpoints to the incident vertices yields the compact metric graph \(|G|\). We denote by \(d_G\) both the resulting shortest-path metric on \(|G|\) and its restriction to \(V\).

Every finite metric space arises as the shortest-path metric of a positively weighted complete graph.

\begin{lemma}[Complete-graph representation of finite metrics]\label{lemma:finite_complete}
Let \((X,d)\) be a finite metric space. Assigning length \(c_{\{x,y\}}=d(x,y)\) to each edge \(\{x,y\}\) of the complete graph \(K_X\) makes the shortest-path metric of \(K_X\) on \(X\) equal to \(d\).
\end{lemma}

\begin{proof}
\renewcommand{\qedsymbol}{\ensuremath{\blacksquare}}
For \(x\ne y\), the edge \(\{x,y\}\) gives \(d_{K_X}(x,y)\le d(x,y)\), whereas by the triangle inequality every path from \(x\) to \(y\) in \(K_X\) has length at least \(d(x,y)\). For \(x=y\), both distances are zero.
\end{proof}

Given metric spaces \((X,d_X)\) and \((Y,d_Y)\), the Lipschitz constant \(\operatorname{Lip}(F)\) of a map \(F:X\to Y\) is the smallest \(L\ge 0\) such that
\[
  d_Y\bigl(F(x),F(x')\bigr) \le L\,d_X(x,x')
\]
for all \(x,x'\in X\), with \(\operatorname{Lip}(F)=+\infty\) if no such \(L\) exists. Maps satisfying \(\operatorname{Lip}(F)\le 1\) are called nonexpansive. A map \(f:X\to Y\) satisfying
\[
  d_X(x,x') \le d_Y\bigl(f(x),f(x')\bigr) \le D\,d_X(x,x')
\]
for all \(x,x'\in X\) is called a noncontractive embedding with distortion at most \(D\).

\subsection{Smooth Surfaces}\label{subsec:smooth_surfaces}

A smooth Riemannian surface \(M\) carries a smoothly varying inner product on the tangent plane at each point, which defines the arc length \(L_M(\gamma)\) of any piecewise smooth curve \(\gamma\) on \(M\). For a smooth surface \(M\subset\mathbb{R}^3\) with the induced metric, measuring tangent vectors with the standard Euclidean inner product of \(\mathbb{R}^3\) gives this arc length directly. When \(M\) is connected, the intrinsic distance between \(x,y\in M\) is defined by
\[
  d_M(x,y) = \inf_{\gamma} L_M(\gamma),
\]
where the infimum is taken over all piecewise smooth curves connecting \(x\) and \(y\) on \(M\).

A compact smooth surface without boundary is called a smooth closed surface; in \(\mathbb{R}^3\), a smooth closed surface is orientable if it admits a continuous unit normal vector field. Topologically, a connected smooth closed orientable surface \(M\) is determined by its number of handles, called its genus \(g(M)\ge 0\). Correspondingly, the orientable genus of a finite undirected graph is the minimum genus of a connected closed orientable surface into which the graph embeds without crossings.

\subsection{Space Information Flow}

Let \((X,d_X)\) be a metric space and let \(\mathcal{D}=\{(s_i,t_i,r_i)\}_{i=1}^k\) be a finite indexed family of independent unicast demands in \(X\), where session \(i\) has source and destination terminals \(s_i,t_i\in X\) with \(s_i\ne t_i\) and target rate \(r_i>0\). Distinct indices denote independent messages even when the terminal pairs and rates coincide. Following~\cite{LiWu2012SIF,XiahouLiWuHuang2014}, a finite scheme network in \(X\) consists of a finite directed multigraph \(N=(W,A)\), a node-placement map \(p:W\to X\), nonnegative link bandwidths \(b_a\ge 0\) (\(a\in A\)), and local routing or network coding operations achieving \(\mathcal{D}\), where \(p\) fixes all terminals at their prescribed locations and places any additional relay or coding nodes arbitrarily in \(X\). We use zero-error network coding. Each node determines its transmissions from its local source messages and earlier receptions, and each destination must recover its requested message exactly. Let \(b_a\) denote the average rate at which the scheme sends data over directed link \(a\). The total cost is
\begin{equation}\label{eq:network_volume}
  \operatorname{Vol}_X(N,p) = \sum_{a\in A} b_a\, d_X\bigl(p(a^-),p(a^+)\bigr),
\end{equation}
where \(a^-\) and \(a^+\) denote the tail and head of arc \(a\); distinct abstract nodes may share the same point in \(X\), and parallel arcs are permitted.

We denote by \(\Route_X(\mathcal{D})\) and \(\Code_X(\mathcal{D})\) the infima of~\eqref{eq:network_volume} over all finite fractional routing scheme networks and finite network coding scheme networks achieving \(\mathcal{D}\) in \(X\), respectively. Similarly, for a finite connected positively weighted undirected graph \(G=(V,E,c)\) and a demand family \(\mathcal{D}\) on \(V\), \(\Route_G(\mathcal{D})\) and \(\Code_G(\mathcal{D})\) denote the optimal fractional routing cost and optimal network coding cost on \(G\), where each undirected edge \(e\in E\) contributes \(c_e\) times the sum of bandwidths transmitted in its two directions.

Given a map \(F:X\to Y\) with \(F(s_i)\ne F(t_i)\) for each session, the mapped demand family is
\begin{equation}\label{eq:demand_map}
  F\mathcal{D} = \bigl\{(F(s_i),F(t_i),r_i)\bigr\}_{i=1}^k.
\end{equation}
Whenever \(\Code_X(\mathcal{D})>0\), the coding advantage is defined by
\begin{equation}\label{eq:advantage}
  \Adv_X(\mathcal{D}) = \frac{\Route_X(\mathcal{D})}{\Code_X(\mathcal{D})},
\end{equation}
with \(\Adv_G(\mathcal{D}) = \Route_G(\mathcal{D})/\Code_G(\mathcal{D})\) defined analogously on graphs.

%% file: sections/main_result.tex
\section{Main Results}\label{sec:main_result}

\subsection{Almost-Isometric Embeddings of Graphs and Finite Metric Spaces}

The vertex set of every finite connected positively weighted graph, equipped with its shortest-path metric, embeds noncontractively into a smooth closed surface in \(\mathbb{R}^3\) with distortion arbitrarily close to \(1\).

\begin{theorem}[Almost-isometric embedding of finite positively weighted graphs]\label{theorem:surface_realization}
Let \(G=(V,E,c)\) be a finite connected undirected simple graph on \(n\) vertices and \(m\) edges with edge lengths \(c_e>0\) for all \(e\in E\). For every \(\varepsilon>0\), there exist a smooth connected closed orientable surface \(M\subset\mathbb{R}^3\) of genus
\begin{equation}\label{eq:genus}
  g(M) = m - n + 1
\end{equation}
and an injective map \(f:V\to M\) such that
\begin{equation}\label{eq:vertex_bounds}
  d_G(u,v) \le d_M\bigl(f(u),f(v)\bigr) \le (1+\varepsilon)\,d_G(u,v), \qquad u,v\in V.
\end{equation}
\end{theorem}

By Lemma~\ref{lemma:finite_complete}, every finite metric space is the shortest-path metric of a positively weighted complete graph and therefore admits an almost-isometric embedding into a smooth closed surface.

\begin{corollary}[Almost-isometric embedding of finite metric spaces]\label{corollary:finite_metric}
Let \((X,d)\) be a finite metric space on \(n\ge 2\) points. For every \(\varepsilon>0\), there exist a smooth connected closed orientable surface \(M\subset\mathbb{R}^3\) of genus
\begin{equation}\label{eq:complete_genus}
  g(M) = \frac{(n-1)(n-2)}{2}
\end{equation}
and an injective map \(f:X\to M\) such that
\begin{equation}\label{eq:finite_bounds}
  d(x,y) \le d_M\bigl(f(x),f(y)\bigr) \le (1+\varepsilon)\,d(x,y), \qquad x,y\in X.
\end{equation}
\end{corollary}

\begin{proof}
\renewcommand{\qedsymbol}{\ensuremath{\blacksquare}}
Apply Theorem~\ref{theorem:surface_realization} to the positively weighted complete graph \(K_X\) of Lemma~\ref{lemma:finite_complete}, which has \(n\) vertices and \(n(n-1)/2\) edges.
\end{proof}

\subsection{Cost and Coding-Advantage Transfer in Space Information Flow}

In space information flow, relay and coding nodes may be placed anywhere on a surface, so controlling distances between graph vertices alone does not bound the total cost of multihop links. For every finite connected undirected simple graph \(G=(V,E,c)\) on \(n\) vertices and \(m\) positively weighted edges and every \(\varepsilon>0\), there exist a smooth connected closed orientable surface \(M\subset\mathbb{R}^3\) of genus \(m-n+1\) and continuous maps (Theorem~\ref{theorem:two_sided_maps} in Section~\ref{sec:surface_realization})
\begin{equation}\label{eq:maps}
  \iota:|G|\to M, \qquad \pi:M\to |G|,
\end{equation}
such that \(\pi(\iota(v))=v\) for all \(v\in V\) and
\begin{equation}\label{eq:lipschitz}
  \operatorname{Lip}(\iota) \le 1+\varepsilon, \qquad \operatorname{Lip}(\pi) \le 1.
\end{equation}
Taking the restriction \(f=\iota|_V\) yields Theorem~\ref{theorem:surface_realization}: the upper bound follows from \(\operatorname{Lip}(\iota)\le 1+\varepsilon\), while \(d_G(u,v)=d_G(\pi(f(u)),\pi(f(v)))\le d_M(f(u),f(v))\) gives the lower bound and injectivity. Moreover, \(\iota\) and \(\pi\) transfer communication schemes in both directions between \(|G|\) and the entire surface \(M\), ensuring that the single surface \(M\) preserves optimal fractional routing cost and optimal network coding cost within a factor of \(1+\varepsilon\) across all finite multiple-unicast demand families on \(V\).

\begin{theorem}[Two-sided preservation of communication costs and coding advantage]\label{theorem:surface_cost}
Let \(G=(V,E,c)\) be a finite connected undirected simple graph on \(n\) vertices and \(m\) positively weighted edges. For every \(\varepsilon>0\), there exist a smooth connected closed orientable surface \(M\subset\mathbb{R}^3\) of genus \(m-n+1\) and an injective map \(f:V\to M\) such that for every finite multiple-unicast demand family \(\mathcal{D}\) on \(V\),
\begin{equation}\label{eq:route_surface}
  \Route_G(\mathcal{D}) \le \Route_M(f\mathcal{D}) \le (1+\varepsilon)\Route_G(\mathcal{D}),
\end{equation}
\begin{equation}\label{eq:code_surface}
  \Code_G(\mathcal{D}) \le \Code_M(f\mathcal{D}) \le (1+\varepsilon)\Code_G(\mathcal{D}).
\end{equation}
Whenever \(\Code_G(\mathcal{D})>0\),
\begin{equation}\label{eq:advantage_transfer}
  \frac{1}{1+\varepsilon}\Adv_G(\mathcal{D}) \le \Adv_M(f\mathcal{D}) \le (1+\varepsilon)\Adv_G(\mathcal{D}).
\end{equation}
\end{theorem}

Theorem~\ref{theorem:surface_cost} transfers the coding advantage of each graph instance to a corresponding surface. Existing counterexamples and the amplification construction give graph instances with arbitrarily large coding advantage. Applying the transfer theorem to these instances gives the following result.

\begin{theorem}[Arbitrarily large coding advantage on smooth closed surfaces]\label{theorem:unbounded_advantage}
For every $A>1$, there exist a smooth connected closed orientable surface $M\subset\mathbb{R}^3$ and a finite multiple-unicast demand family $\mathcal{D}$ on $M$ such that
\begin{equation}\label{eq:unbounded_advantage}
  \Route_M(\mathcal{D})>A\,\Code_M(\mathcal{D})>0.
\end{equation}
\end{theorem}

\subsection{Genus Upper Bound on Coding Advantage}

Whenever the orientable genus of a graph or the genus of a smooth closed surface is at most \(g\), multiple-unicast coding advantage is bounded uniformly by \(O(\log(g+2))\).

\begin{theorem}[Genus upper bound on coding advantage]\label{theorem:genus_bound}
There exists an absolute constant \(C>0\) such that for every integer \(g\ge 0\), if \(X\) is either a finite connected positively weighted undirected graph of orientable genus at most \(g\) or a smooth connected closed orientable Riemannian surface of genus at most \(g\), then every finite multiple-unicast demand family \(\mathcal{D}\) on \(X\) satisfies
\begin{equation}\label{eq:genus_advantage_bound}
  \Route_X(\mathcal{D}) \le C\log(g+2)\,\Code_X(\mathcal{D}),
\end{equation}
where \(\log\) denotes the natural logarithm. Setting \(g=0\) shows that multiple-unicast coding advantage on planar graphs and on smooth closed surfaces of genus \(0\) (topological spheres) is bounded by an absolute constant. Moreover, if \(\Code_X(\mathcal{D})>0\) and \(\Adv_X(\mathcal{D})\ge A\), then \(g+2 \ge \exp(A/C)\).
\end{theorem}

%% file: sections/proof_overview.tex
\section{Proof Overview}\label{sec:proof_overview}

The core of Theorems~\ref{theorem:surface_realization} and~\ref{theorem:surface_cost} is constructing the two-sided maps \(\iota:|G|\to M\) and \(\pi:M\to|G|\) in~\eqref{eq:maps}--\eqref{eq:lipschitz} between the metric graph \(|G|\) and a smooth closed surface \(M\subset\mathbb{R}^3\). We first draw \(G\) in \(\mathbb{R}^3\) without crossings so that each edge has a prescribed spatial arc length. Because arbitrary positive edge lengths cannot in general be drawn directly using straight-line segments in \(\mathbb{R}^3\) without crossings, we first place the vertices at distinct points along the curve \((t,t^2,t^3)\), where no three points are collinear and no four are coplanar, ensuring that straight segments between adjacent vertices never cross; scaling the drawing down makes every segment shorter than its target length. Around the middle of each edge we then place pairwise disjoint thin solid cylinders and replace the straight axial segment inside each cylinder with a smoothly oscillating curve. Choosing a sufficiently high oscillation frequency and then adjusting the amplitude lengthens each edge independently to any desired value without touching other edges or altering the straight directions near the vertices.

Next, we thicken this spatial graph into a smooth solid tubular region of small radius \(r\) and take its boundary as our smooth closed orientable surface \(M\). Topologically, thickening the graph turns each independent cycle into a handle, giving \(M\) genus \(m-n+1\). Geometrically, \(M\) consists of small vertex junctions connected by thin circular tubes around the edges: collapsing each vertex junction to its graph vertex and projecting each circular cross-section of a tube onto the corresponding point of the graph edge defines \(\pi:M\to|G|\), while choosing a representative point in each vertex junction and connecting adjacent representatives by curves along the tubes and through their end junctions defines \(\iota:|G|\to M\). A key detail arises here: the inner side of a curved tube is slightly shorter than its core curve, and small segments at both ends of each edge are absorbed into the vertex junctions; had we set the spatial edge lengths equal to \(c_e\), a path on the tube wall could be shorter than its projection onto the graph edge, causing \(\pi\) to stretch distances. We therefore fix \(0<\eta<\varepsilon\) in the first step and set each target spatial edge length to \(L_e=(1+\eta)c_e\). Once the tube radius \(r\) is sufficiently small, junction diameters, end truncations, and curvature-induced length differences all become negligible compared with the extra length \(\eta c_e\), ensuring that \(\pi\) never increases distances while \(\iota\) stretches distances by at most \(1+\varepsilon\). The three stages are illustrated in Fig.~\ref{fig:three_steps}.

\mfInsertFigure{construction}{\mfConstructionCaption}{fig:three_steps}

With \(\iota\) and \(\pi\) in hand, communication schemes transfer in both directions between the discrete graph \(G\) and the smooth closed surface \(M\), allowing us to compare the optimal fractional routing and network coding costs in the two settings. A graph edge costs its bandwidth times its length, and a surface link costs its bandwidth times the intrinsic distance between its endpoints. Mapping the vertices of any routing or coding scheme on \(G\) to \(M\) via \(\iota\) while keeping all bandwidths and node operations unchanged gives each link a length at most \(1+\varepsilon\) times the original edge length, so optimal costs on \(M\) are at most \(1+\varepsilon\) times those on \(G\). Conversely, even if a finite scheme on \(M\) places relay and coding nodes anywhere on the surface, the nonexpansive map \(\pi\) projects all nodes onto vertices or edge interiors of \(|G|\), returning every terminal to its original vertex in \(V\) without increasing any link distance. After each projected link is expanded along a shortest path, nodes landing inside an edge of \(|G|\) split that edge into segments in series; since no demand terminals lie inside the edge, replacing the segmented edge by a single edge of capacity equal to the minimum aggregate bidirectional bandwidth among its segments preserves end-to-end communication without increasing cost, showing that optimal costs on \(M\) are at least those on \(G\). Dividing the bounds for the two optimal costs gives the two-sided estimate for coding advantage in Theorem~\ref{theorem:surface_cost}. For fixed $\varepsilon>0$, if the coding advantages of a sequence of graph instances tend to infinity, so do those of the corresponding surface instances. This proves Theorem~\ref{theorem:unbounded_advantage}.

Finally, we prove the \(O(\log(g+2))\) coding-advantage bound at fixed genus \(g\) in Theorem~\ref{theorem:genus_bound}. For a finite graph of orientable genus at most \(g\), combining constant-distortion \(L_1\) embeddings of planar graphs with stochastic planarization embeds the graph's shortest-path metric into \(L_1\) with distortion \(O(\log(g+2))\). Every finite \(L_1\) metric is a nonnegative weighted sum of cut metrics; across each cut separating the vertices into two sides, any valid coding scheme must carry at least as much bidirectional traffic as the total demand separated by that cut. Summing these cut inequalities with the \(L_1\) weights bounds optimal fractional routing cost by \(O(\log(g+2))\) times optimal coding cost. For a smooth closed orientable surface of genus at most \(g\), we draw the links of any finite coding scheme as minimizing geodesics on the surface and perturb them slightly so that they cross only at finitely many interior points with arbitrarily small cost increase. Inserting a pure forwarding relay at every crossing turns the surface scheme into a finite graph drawn without crossings on the surface; each connected component has orientable genus at most \(g\), so applying the bounded-genus graph bound yields the same bound on the surface.

%% file: sections/spatial_embeddings.tex
\section{Spatial Embeddings with Prescribed Edge Lengths}\label{sec:spatial_embeddings}

\begin{theorem}[Spatial embedding with prescribed edge lengths]\label{theorem:spatial_embedding}
Let \(G=(V,E)\) be a finite undirected simple graph with prescribed positive edge lengths \(L_e>0\) for each \(e\in E\). Then \(G\) admits a spatial embedding \(\Gamma\subset\mathbb{R}^3\) in which each edge is a smooth embedded arc of length \(L_e\) that is a straight-line segment near both of its endpoints, distinct edges intersect only at common endpoints, and edges incident to a common vertex have distinct tangent directions at that vertex.
\end{theorem}

\begin{proof}[Proof of Theorem~\ref{theorem:spatial_embedding}]
\renewcommand{\qedsymbol}{\ensuremath{\blacksquare}}
Choose distinct real numbers \(t_v\) for \(v\in V\) and place the vertices at
\begin{equation}\label{eq:moment_curve}
  p_v = (t_v, t_v^2, t_v^3) \in \mathbb{R}^3.
\end{equation}
By the Vandermonde determinant, no three of these points are collinear and no four are coplanar~\cite[Lemma~5.4.2]{Matousek2002}. Consequently, connecting pairs of vertices by straight-line segments yields a crossing-free embedding of the complete graph in \(\mathbb{R}^3\): two segments without a common endpoint can intersect only if their four endpoints are coplanar, and two segments sharing an endpoint can overlap only if their three endpoints are collinear. Restricting this straight-line drawing to \(G\) and applying a uniform scaling of \(\mathbb{R}^3\) produces a crossing-free straight-line embedding \(\Gamma_0\subset\mathbb{R}^3\) in which incident edges have distinct directions at every vertex and the straight segment \(\Gamma_{0,e}\) representing each edge \(e\in E\) has length \(l_e^0<L_e\).

For each edge \(e\in E\), pick an interior point \(x_e\) of the segment \(\Gamma_{0,e}\) and an open subsegment \(I_e\) of \(\Gamma_{0,e}\) containing \(x_e\). Because the compact set \(\Gamma_0\setminus I_e\) does not contain \(x_e\), it lies at a positive Euclidean distance from \(x_e\), so there is an open ball \(B_e\subset\mathbb{R}^3\) centered at \(x_e\) with \(B_e\cap\Gamma_0\subset I_e\). Since \(E\) is finite and the points \(x_e\) are distinct, we may shrink the balls \(B_e\) so that their closures are pairwise disjoint. Inside each \(B_e\), choose a closed solid right circular cylinder \(Z_e\subset B_e\) whose axis \(A_e\) is a closed subsegment of \(B_e\cap\Gamma_{0,e}\). The cylinders \(Z_e\) are then pairwise disjoint and satisfy
\begin{equation}\label{eq:cylinder_axis}
  Z_e \cap \Gamma_0 = A_e.
\end{equation}

Fix an edge \(e\in E\). In local Euclidean coordinates, write the cylinder axis as
\begin{equation}\label{eq:axis_interval}
  A_e = \{(t,0,0) : -a_e \le t \le a_e\}
\end{equation}
with \(a_e>0\). Choose a nonnegative smooth function \(\psi_e:[-a_e,a_e]\to\mathbb{R}_{\ge 0}\) that is not identically zero on \((-a_e,a_e)\) and vanishes identically in neighborhoods of \(-a_e\) and \(a_e\), and consider replacing \(A_e\) by the curve
\begin{equation}\label{eq:oscillation}
  \gamma_{\alpha,N}(t) = \bigl(t,\,\alpha\psi_e(t)\sin(Nt),\,0\bigr), \qquad -a_e \le t \le a_e,
\end{equation}
for parameters \(\alpha\ge 0\) and \(N>0\). Because its first coordinate is \(t\), \(\gamma_{\alpha,N}\) is a smooth embedded arc; because \(\psi_e\) vanishes near \(\pm a_e\), the arc coincides with \(A_e\) near both ends and therefore joins smoothly with the remainder of \(\Gamma_{0,e}\).

Let \(M_e=\max_{t\in[-a_e,a_e]}\psi_e(t)>0\), and choose \(\alpha_0>0\) small enough that \(\alpha_0 M_e\) is strictly less than the radius of \(Z_e\). Then \(\gamma_{\alpha,N}\) lies entirely inside \(Z_e\) for all \(0\le\alpha\le\alpha_0\) and all \(N>0\). When \(\alpha=\alpha_0\), the arc length of \(\gamma_{\alpha_0,N}\) tends to infinity as \(N\to\infty\). Indeed, choose a closed subinterval \(J\subset(-a_e,a_e)\) of positive length and a constant \(\mu_e>0\) such that \(\psi_e(t)\ge\mu_e\) on \(J\). At the points
\begin{equation}\label{eq:alternating_points}
  t_k = \frac{\pi/2+k\pi}{N}, \qquad k\in\mathbb{Z},
\end{equation}
lying in \(J\), the second coordinate of \(\gamma_{\alpha_0,N}(t_k)\) alternates in sign with magnitude at least \(\alpha_0\mu_e\), so consecutive points \(\gamma_{\alpha_0,N}(t_k)\) and \(\gamma_{\alpha_0,N}(t_{k+1})\) are at Euclidean distance at least \(2\alpha_0\mu_e\). Since the number \(N_J\) of such points in \(J\) satisfies \(N_J\ge N|J|/\pi-2\), the total arc length of \(\gamma_{\alpha_0,N}\) is at least \(2\alpha_0\mu_e(N_J-1)\), which diverges to infinity as \(N\to\infty\).

Fix \(N\) large enough that replacing \(A_e\) by \(\gamma_{\alpha_0,N}\) makes the total length of edge \(e\) strictly greater than \(L_e\). With \(N\) fixed, the total length of edge \(e\) is a continuous function of \(\alpha\in[0,\alpha_0]\) that equals \(l_e^0<L_e\) at \(\alpha=0\) and exceeds \(L_e\) at \(\alpha=\alpha_0\). By the intermediate value theorem, some \(\alpha\in(0,\alpha_0)\) gives edge \(e\) exact total length \(L_e\).

Performing this replacement independently inside each cylinder \(Z_e\) (\(e\in E\)) introduces no self-intersections or intersections between distinct edges, because the cylinders \(Z_e\) are pairwise disjoint and each meets \(\Gamma_0\) only along \(A_e\). The straight segments and distinct tangent directions near every vertex remain unchanged, completing the construction of \(\Gamma\subset\mathbb{R}^3\).
\end{proof}

%% file: sections/surface_realization.tex
\section{Construction of the Smooth Surface}\label{sec:surface_realization}

We thicken the spatially embedded graph \(\Gamma\subset\mathbb{R}^3\) constructed in Section~\ref{sec:spatial_embeddings} to obtain a smooth closed orientable surface \(M_r\subset\mathbb{R}^3\). We construct the two maps \(\iota:|G|\to M_r\) and \(\pi:M_r\to|G|\) in~\eqref{eq:maps}--\eqref{eq:lipschitz} between the metric graph \(|G|\) and \(M_r\), thereby proving Theorem~\ref{theorem:surface_realization}.

Let \(G=(V,E)\) be a finite connected undirected simple graph with \(n\) vertices and \(m\ge 1\) edges, and let \(\Gamma\subset\mathbb{R}^3\) be the spatial embedding constructed in Theorem~\ref{theorem:spatial_embedding} for positive edge lengths \(L_e>0\). Near each vertex position \(p_v\in\mathbb{R}^3\), the incident edges are straight segments with distinct directions. For each vertex \(v\), fix a connected compact smooth template surface \(U_v^0\subset\mathbb{R}^3\) at the origin. Starting from a sphere of sufficiently large radius centered at the origin, open a hole in each incident edge direction and smoothly extend each opening into a short straight cylindrical outlet of radius \(1\). The outlets are disjoint and together form a junction for the incident tubes. For \(r>0\), scale the template by \(r\) and translate it to \(p_v\) to define the vertex junction
\begin{equation}\label{eq:scaled_junction}
  U_v=p_v+rU_v^0.
\end{equation}
For sufficiently small \(r\), each outlet of \(U_v\) is a straight cylindrical surface of radius \(r\) around the corresponding straight edge segment near \(p_v\). By the tubular neighborhood construction~\cite{Hirsch1976,FriedlHerrmann2021}, there is an \(r_0>0\) such that for all \(0<r<r_0\), trimming both ends of each edge leaves a core arc \(\gamma_e\) of length \(\ell_e\) around which a circular tube \(T_e\) of radius \(r\) smoothly joins the two junction outlets. The pieces form a smooth embedded connected closed orientable surface \(M_r\) in \(\mathbb{R}^3\). Each independent cycle of \(G\) gives rise to a handle in the thickening, so the genus of \(M_r\) is
\begin{equation}\label{eq:euler}
  g(M_r)=m-n+1.
\end{equation}
The junctions \(U_v\) and tubes \(T_e\) have pairwise disjoint interiors. A junction and a tube meet along a common boundary circle exactly when \(v\) is incident to \(e\). Write \(\|\cdot\|\) and \(\langle\cdot,\cdot\rangle\) for the standard Euclidean norm and inner product on \(\mathbb{R}^3\). The length of a piecewise smooth curve \(X:[a,b]\to M_r\) is \(L_{M_r}(X)=\int_a^b\|X'(t)\|\,dt\). Parametrize each core arc by arclength as \(\gamma_e:[0,\ell_e]\to\mathbb{R}^3\), so that \(\|\gamma_e'(s)\|=1\). For sufficiently small \(r_0\), every point of \(T_e\) has a unique representation, depending smoothly on the point, of the form
\begin{equation}\label{eq:parametrization}
  x=\gamma_e(s)+ru,\qquad s\in[0,\ell_e],\quad\|u\|=1,\quad\langle u,\gamma_e'(s)\rangle=0.
\end{equation}

A vertex junction cuts off a short straight segment from the end of each incident edge, and crossing the junction itself adds a short path. Since \(U_v\) is obtained by scaling a fixed template by \(r\), both lengths are \(O(r)\).

\begin{lemma}[Length bounds for junctions and core arcs]\label{lemma:junction_core_lengths}
There are constants \(A,C>0\), independent of \(r\), such that for all \(0<r<r_0\), \(v\in V\), and \(e\in E\), we have \(L_e-Ar\le\ell_e\le L_e\), and any two points of \(U_v\) can be joined within \(U_v\) by a piecewise smooth curve of length at most \(Cr\).
\end{lemma}

\begin{proof}
Let \(a_{v,e}\) be the axial distance from the origin to the center of the end circle of the outlet of \(U_v^0\) in the direction of edge \(e\). The straight segments trimmed from the two ends of \(e=\{u,v\}\) then have total length \((a_{u,e}+a_{v,e})r\). Taking \(A=\max_{e=\{u,v\}\in E}(a_{u,e}+a_{v,e})\) gives \(L_e-Ar\le\ell_e\le L_e\). Each template \(U_v^0\) is a connected compact smooth surface and has finite intrinsic diameter. Choose \(C\) larger than the maximum of these diameters. Any two points of a template can be joined within it by a piecewise smooth curve of length at most \(C\). Scaling by \(r\) and translating to \(U_v\) multiplies this length by \(r\), giving the junction bound.
\end{proof}

To define the projection \(\pi\), we will rescale the axial coordinate \(s\) of a tube to a coordinate on the graph edge. Along the inside of a bend, the axial coordinate can change slightly faster than arclength. The next lemma bounds this factor by \(1+O(r)\).

\begin{lemma}[Axial displacement on a tube]\label{lemma:tube_axial_displacement}
There is a constant \(K\ge 0\), independent of \(r\), such that whenever \(0<r<r_0\) and \(rK\le 1/2\), every piecewise smooth curve \(X(t)=\gamma_e(s(t))+ru(t)\), \(t\in[a,b]\), on any tube \(T_e\) satisfies
\begin{equation}\label{eq:longitudinal_bound}
  |s(b)-s(a)|\le(1+2Kr)L_{M_r}(X).
\end{equation}
\end{lemma}

\begin{proof}
The finitely many smooth edge arcs of \(\Gamma\) are fixed. Let \(K\ge 0\) be their maximum curvature before trimming, so that \(\|\gamma_e''(s)\|\le K\) on every trimmed core arc. On each smooth subinterval, differentiating \(\langle u(t),\gamma_e'(s(t))\rangle=0\) gives
\[
  \langle u'(t),\gamma_e'(s(t))\rangle=-s'(t)\langle u(t),\gamma_e''(s(t))\rangle.
\]
Taking the inner product of \(X'(t)=s'(t)\gamma_e'(s(t))+ru'(t)\) with the unit vector \(\gamma_e'(s(t))\) and using the Cauchy--Schwarz inequality and \(\|u(t)\|=1\), we obtain
\begin{equation}\label{eq:longitudinal_derivative}
  \|X'(t)\|\ge\bigl|\langle X'(t),\gamma_e'(s(t))\rangle\bigr|
  =|s'(t)|\,\bigl|1-r\langle u(t),\gamma_e''(s(t))\rangle\bigr|
  \ge(1-rK)|s'(t)|.
\end{equation}
For \(rK\le 1/2\) we have \((1-rK)^{-1}\le 1+2Kr\), hence \(|s'(t)|\le(1+2Kr)\|X'(t)\|\). Integrating over \([a,b]\) yields~\eqref{eq:longitudinal_bound}.
\end{proof}

In the other direction, to construct \(\iota:|G|\to M_r\), we must connect representative points in the two end junctions of each tube. Each spatial edge constructed in the proof of Theorem~\ref{theorem:spatial_embedding} lies in its own plane. Translating its core arc by \(r\) along a normal to that plane gives a curve of the same length on the tube. Adding short paths in the two junctions gives total length at most \(L_e+O(r)\).

\begin{lemma}[Length of a curve connecting the end junctions]\label{lemma:tube_connectors}
Choose an arbitrary point \(q_v\in U_v\) in each junction. There is a constant \(B>0\), independent of \(r\), such that for all \(0<r<r_0\) and each edge \(e=\{u,v\}\in E\), there is a piecewise smooth curve \(\lambda_e\) on \(M_r\) joining \(q_u\) to \(q_v\) with \(L_{M_r}(\lambda_e)\le L_e+Br\).
\end{lemma}

\begin{proof}
In the local Euclidean coordinates of~\eqref{eq:oscillation}, the oscillating portion and the straight segments at both ends lie in the plane with third coordinate zero. Choose a constant unit normal \(w_e\) to that plane. Then \(\langle w_e,\gamma_e'(s)\rangle=0\) for all \(s\in[0,\ell_e]\), so the translated curve \(X_e(s)=\gamma_e(s)+rw_e\), \(s\in[0,\ell_e]\), lies entirely on \(T_e\). Its length is exactly \(\ell_e\le L_e\) because \(w_e\) is constant. By Lemma~\ref{lemma:junction_core_lengths}, we can join \(q_u\) to \(X_e(0)\) in \(U_u\) and \(X_e(\ell_e)\) to \(q_v\) in \(U_v\) by piecewise smooth curves of length at most \(Cr\) each. Concatenating the three curves gives \(\lambda_e\), and taking \(B=2C\) gives \(L_{M_r}(\lambda_e)\le L_e+Br\).
\end{proof}

We now define the projection \(\pi:M_r\to|G|\). Map each junction \(U_v\) to the vertex \(v\) of \(|G|\). For each edge \(e=\{u,v\}\in E\), orient the core from \(u\) to \(v\) so that \(s=0\) meets \(U_u\) and \(s=\ell_e\) meets \(U_v\). Map a point of \(T_e\) with axial coordinate \(s\) to the point at arclength \(c_e s/\ell_e\) from \(u\) along \(e\), where \(c_e\) is the length of \(e\) in the metric graph. The definitions agree on the common boundary circles, so \(\pi\) is continuous on \(M_r\). When the core length \(\ell_e\) absorbs the factor \(1+2Kr\) from Lemma~\ref{lemma:tube_axial_displacement}, we can show that \(\pi\) does not increase distances by subdividing curves into subarcs in tube interiors, junction interiors, or straight cylindrical neighborhoods of the attachment circles. Figure~\ref{fig:tube_maps} shows the local tube geometry and both maps.

\mfInsertFigure{tube-maps}{\mfTubeCaption}{fig:tube_maps}

\begin{lemma}[Nonexpansiveness of \(\pi\)]\label{lemma:nonexpansive_collapse}
If \(0<r<r_0\) satisfies \(rK\le 1/2\) and \(c_e(1+2Kr)\le\ell_e\) for every \(e\in E\), then \(\operatorname{Lip}(\pi)\le 1\).
\end{lemma}

\begin{proof}
In a small neighborhood on both sides of each boundary circle \(U_v\cap T_e\), the surface is a straight circular cylinder of radius \(r\) around the straight edge segment. Let \(e_v\) be the unit vector along the segment pointing into \(T_e\). Fix a point \(x_0\) on the boundary circle and write \(h(x)=\langle x-x_0,e_v\rangle\) for the axial coordinate. Then \(h\) is nonnegative on the tube side and nonpositive on the junction side, and \(\pi(x)\) lies at arclength \((c_e/\ell_e)\max\{h(x),0\}\) from \(v\) along \(e\). Let \(X:[a,b]\to M_r\) be any piecewise smooth curve joining \(x\) and \(y\). The tube interiors, junction interiors, and straight cylindrical neighborhoods cover \(X([a,b])\). By continuity of \(X\) and compactness of \([a,b]\), we can subdivide \([a,b]\) into finitely many closed intervals \([t_{j-1},t_j]\) such that each subarc \(X_j=X|_{[t_{j-1},t_j]}\) lies entirely in one of these open sets.
\begin{enumerate}
\item In the interior of \(T_e\), Lemma~\ref{lemma:tube_axial_displacement} and \(c_e(1+2Kr)\le\ell_e\) bound the distance in \(|G|\) between the endpoint images by \((c_e/\ell_e)|s(t_j)-s(t_{j-1})|\le L_{M_r}(X_j)\).
\item In the interior of \(U_v\), the map \(\pi\) is constant with value \(v\), so the endpoint images have distance zero.
\item In a straight cylindrical neighborhood, the Cauchy--Schwarz inequality gives \(|(h\circ X)'(\tau)|\le\|X'(\tau)\|\). Taking the positive part is nonexpansive, and \(c_e/\ell_e\le1\), so the distance between the endpoint images is at most \(|h(X(t_j))-h(X(t_{j-1}))|\le L_{M_r}(X_j)\).
\end{enumerate}
Summing over the subarcs and applying the triangle inequality for \(d_G\) gives \(d_G(\pi(x),\pi(y))\le L_{M_r}(X)\). Taking the infimum over all piecewise smooth curves from \(x\) to \(y\) gives \(d_G(\pi(x),\pi(y))\le d_{M_r}(x,y)\).
\end{proof}

Finally, prescribe the spatial edge lengths \(L_e=(1+\eta)c_e\) with \(0<\eta<\varepsilon\). For sufficiently small tube radius \(r\), the \(O(r)\) errors in Lemmas~\ref{lemma:junction_core_lengths}--\ref{lemma:tube_connectors} allow us to meet the nonexpansiveness condition of Lemma~\ref{lemma:nonexpansive_collapse} while keeping each connecting curve \(\lambda_e\) of length at most \((1+\varepsilon)c_e\).

\begin{theorem}[Two-sided maps between metric graphs and smooth closed surfaces]\label{theorem:two_sided_maps}
Let \(G=(V,E,c)\) be a finite connected undirected simple graph with \(n\) vertices and \(m\) positively weighted edges. For any \(\varepsilon>0\), there exist a smooth connected closed orientable surface \(M\subset\mathbb{R}^3\) of genus \(m-n+1\) and continuous maps \(\iota:|G|\to M\) and \(\pi:M\to|G|\) such that \(\operatorname{Lip}(\iota)\le 1+\varepsilon\), \(\operatorname{Lip}(\pi)\le 1\), and \(\pi(\iota(v))=v\) for all \(v\in V\).
\end{theorem}

\begin{proof}[Proof of Theorems~\ref{theorem:two_sided_maps} and~\ref{theorem:surface_realization}]
If \(n=1\), then \(m=0\), and we take the unit sphere in \(\mathbb{R}^3\) and constant maps. Suppose \(m\ge1\). Fix \(0<\eta<\varepsilon\), set \(L_e=(1+\eta)c_e\), and construct the spatial embedding \(\Gamma\subset\mathbb{R}^3\) by Theorem~\ref{theorem:spatial_embedding}. As \(r\to0\), we have \(c_e(1+2Kr)/(L_e-Ar)\to(1+\eta)^{-1}<1\) and \((L_e+Br)/c_e\to1+\eta<1+\varepsilon\) for every edge \(e\in E\). Since \(E\) is finite, we can choose \(0<r<r_0\) with \(rK\le1/2\) and \(Ar<\min_{e\in E}L_e\) such that simultaneously for all \(e\in E\),
\begin{equation}\label{eq:collapse_choice}
  c_e(1+2Kr)\le L_e-Ar\le\ell_e,\qquad L_e+Br\le(1+\varepsilon)c_e.
\end{equation}
Lemma~\ref{lemma:nonexpansive_collapse} gives \(\operatorname{Lip}(\pi)\le1\). Set \(\iota(v)=q_v\in U_v\), and map the interval of length \(c_e\) representing each edge \(e=\{u,v\}\) of \(|G|\) at constant speed along the curve \(\lambda_e\subset M_r\) supplied by Lemma~\ref{lemma:tube_connectors}. These maps agree at the common vertices and give a continuous map \(\iota:|G|\to M_r\). By Lemma~\ref{lemma:tube_connectors} and the last inequality in~\eqref{eq:collapse_choice}, \(L_{M_r}(\lambda_e)\le(1+\varepsilon)c_e\). The shortest-path metric \(d_G\) therefore gives \(\operatorname{Lip}(\iota)\le1+\varepsilon\), and \(q_v\in U_v\) gives \(\pi(\iota(v))=v\) for every \(v\in V\). Setting \(M=M_r\) proves Theorem~\ref{theorem:two_sided_maps}. Restricting to the vertices gives \(f=\iota|_V\) with \(d_G(u,v)=d_G(\pi(f(u)),\pi(f(v)))\le d_M(f(u),f(v))\le(1+\varepsilon)d_G(u,v)\), proving Theorem~\ref{theorem:surface_realization}.
\end{proof}

%% file: sections/space_information_flow.tex
\section{Costs and Coding Advantage in Space Information Flow}\label{sec:space_information_flow}

Two observations allow us to compare communication costs on a discrete graph \(G\) and a smooth closed surface \(M\). First, moving all nodes of a space information flow scheme by a Lipschitz map increases its cost by at most the Lipschitz factor. Second, when the projection of a surface scheme places nodes inside the edges of the metric graph \(|G|\), the relay and coding nodes introduced by subdividing those edges cannot lower optimal costs on the original discrete graph \(G\). We establish these facts, compare costs in both directions between graphs and surfaces, and then bound coding advantage at fixed genus.

\subsection{Optimal Routing Cost and Cost Transfer under Maps}

In fractional routing for multiple unicast sessions, the sessions remain separate. Each session can use a direct link between its terminals in a metric space or a shortest path in a graph. Thus, on any metric space or positively weighted undirected graph, optimal fractional routing cost is the rate-weighted sum of the source--sink distances.

\begin{lemma}[Optimal fractional routing cost formula]\label{lemma:routing_cost}
For any metric space \((X,d_X)\) with a finite multiple-unicast demand family \(\mathcal{D}=\{(s_i,t_i,r_i)\}_{i=1}^k\), and for any finite connected positively weighted undirected graph \(G=(V,E,c)\) with demands \(\mathcal{D}\) on \(V\), the respective costs are

\begin{equation}\label{eq:routing_formula}
  \Route_X(\mathcal{D}) = \sum_{i=1}^k r_i d_X(s_i,t_i), \qquad \Route_G(\mathcal{D}) = \sum_{i=1}^k r_i d_G(s_i,t_i).
\end{equation}
\end{lemma}

\begin{proof}
In \((X,d_X)\), a direct arc of bandwidth \(r_i\) from \(s_i\) to \(t_i\) for each session \(i\) attains the cost on the right of the first equality in~\eqref{eq:routing_formula}. On \(G\), sending flow \(r_i\) along a shortest path from \(s_i\) to \(t_i\) attains the second cost. Conversely, after removing directed cycles from a feasible fractional routing scheme, flow decomposition expresses the flow of each session \(i\) as a sum of directed \(s_i\)--\(t_i\) path flows with total rate \(r_i\). By the triangle inequality or the definition of graph distance, every such path has length at least \(d_X(s_i,t_i)\) or \(d_G(s_i,t_i)\), respectively. Weighting the path lengths by their rates, summing over all sessions, and taking the infimum over feasible routings gives the reverse inequality.
\end{proof}

Network coding has no comparable explicit cost formula in the space information flow model, but the cost of each link is proportional to the distance between its endpoints. Moving all nodes of a scheme to another metric space by an \(L\)-Lipschitz map while retaining their local operations therefore increases both routing and coding costs by at most a factor of \(L\).

\begin{lemma}[Cost transfer under Lipschitz maps]\label{lemma:lipschitz_transport}
Suppose \(F:X\to Y\) satisfies \(\operatorname{Lip}(F)\le L<\infty\) and is injective on the set of terminal locations in \(\mathcal{D}\). Then

\begin{equation}\label{eq:lipschitz_transport}
  \Route_Y(F\mathcal{D}) \le L\,\Route_X(\mathcal{D}), \qquad \Code_Y(F\mathcal{D}) \le L\,\Code_X(\mathcal{D}).
\end{equation}
\end{lemma}

\begin{proof}
Take any finite routing or network coding scheme \((N,p)\) meeting \(\mathcal{D}\) in \(X\). Keep the directed multigraph \(N\), the link bandwidths \(b_a\), and all local routing or coding operations, and replace the placement \(p\) by \(F\circ p\). Since \(F\) is injective on the terminal locations, the resulting scheme meets \(F\mathcal{D}\) in \(Y\). By the Lipschitz bound, the cost of each arc \(a\) in~\eqref{eq:network_volume} satisfies \(b_a d_Y(F(p(a^-)),F(p(a^+)))\le L\,b_a d_X(p(a^-),p(a^+))\). Summing over all arcs and taking the infimum over feasible schemes in \(X\) gives~\eqref{eq:lipschitz_transport}.
\end{proof}

\subsection{Transfer of Costs and Coding Advantage between Graphs and Smooth Closed Surfaces}

The two maps of Theorem~\ref{theorem:two_sided_maps} connect the compact metric graph \(|G|\) and the smooth closed surface \(M\). When \(\pi:M\to|G|\) projects a communication scheme on \(M\) back to \(|G|\), its relay and coding nodes may fall in edge interiors, dividing an edge of the discrete graph \(G\) into a series of segments. To return to \(G\), we use the link expansion and path contraction operations of Li and Li~\cite{LiLi2004} to show that introducing subdivision nodes inside graph edges preserves both optimal routing cost and optimal coding cost.

\begin{lemma}[Cost invariance under edge subdivision]\label{lemma:subdivision}
Let \(G'\) be obtained by finitely subdividing the edges of a finite connected positively weighted undirected graph \(G=(V,E,c)\). Suppose all segment lengths are positive and the lengths on each subdivided edge \(e\) sum to \(c_e\). For every finite multiple-unicast demand family \(\mathcal{D}\) on \(V\),

\begin{equation}\label{eq:subdivision}
  \Route_{G'}(\mathcal{D}) = \Route_G(\mathcal{D}), \qquad \Code_{G'}(\mathcal{D}) = \Code_G(\mathcal{D}).
\end{equation}
\end{lemma}

\begin{proof}
Take any feasible scheme on \(G\). On each original edge \(e\), forward its transmissions along the corresponding subdivided path in \(G'\) with the same aggregate bandwidth \(b_e\) in the two directions. The scheme remains feasible, and the cost on the path is \(b_e\sum_{j=1}^q a_j=b_e c_e\), where \(a_j\) are the segment lengths. Thus optimal costs on \(G'\) are no greater than those on \(G\).

Conversely, take any feasible routing or coding scheme on \(G'\) meeting \(\mathcal{D}\). Suppose an original edge \(e\in E\) has been divided into \(q\) segments of lengths \(a_1,\dots,a_q\), with aggregate bandwidths \(b_1,\dots,b_q\) in the two directions. All terminals lie in \(V\), so the interior of this path contains neither a branch nor a demand terminal. The path contraction of Li and Li~\cite{LiLi2004} replaces the whole path, including its internal forwarding or coding operations, by the original edge \(e\) of capacity \(\min_{1\le j\le q} b_j\) while preserving end-to-end rate feasibility for all sessions. The cost of the contracted edge is

\begin{equation}\label{eq:contracted_cost}
  c_e \min_{1\le j\le q} b_j = \Bigl(\sum_{j=1}^q a_j\Bigr)\min_{1\le j\le q} b_j \le \sum_{j=1}^q a_j b_j.
\end{equation}

Contracting every subdivided edge gives a feasible scheme on \(G\) of no greater cost, proving~\eqref{eq:subdivision}.
\end{proof}

In the space information flow model on the compact metric graph \(|G|\), an abstract link between two points is charged according to their shortest-path distance \(d_G\). Expanding each such link along a shortest path in \(|G|\) gives a path in a finite subdivision, so Lemma~\ref{lemma:subdivision} implies that optimal costs on the discrete graph \(G\) and the metric graph \(|G|\) are equal.

\begin{lemma}[Equality of optimal costs on \(G\) and \(|G|\)]\label{lemma:metric_graph}
For every finite connected positively weighted undirected graph \(G=(V,E,c)\) and every finite multiple-unicast demand family \(\mathcal{D}\) on \(V\),

\begin{equation}\label{eq:graph_equality}
  \Route_{|G|}(\mathcal{D}) = \Route_G(\mathcal{D}), \qquad \Code_{|G|}(\mathcal{D}) = \Code_G(\mathcal{D}).
\end{equation}
\end{lemma}

\begin{proof}
Place any feasible scheme on \(G\) at the same vertices in \(|G|\). Bandwidth used in either direction on an edge \(e=\{u,v\}\) is now charged according to the shortest-path distance \(d_G(u,v)\le c_e\). For example, the endpoints of the length-three edge in a triangle with edge lengths \(1,1,3\) are at distance two in \(|G|\). The total cost cannot increase, so the optimal routing and coding costs on \(|G|\) are no greater than their respective optima on \(G\).

Conversely, take a finite scheme \((N,p)\) on \(|G|\) meeting \(\mathcal{D}\). For each arc \(a\in A\), choose a shortest path in \(|G|\) from \(p(a^-)\) to \(p(a^+)\). Introduce forwarding nodes at the intervening vertices to carry the same message with bandwidth \(b_a\), preserving the operations of any abstract nodes already at those positions. The scheme remains feasible, and the total cost stays the same because the path length is \(d_G(p(a^-),p(a^+))\). After expanding all links, the original vertices \(V\) and the finitely many abstract node positions in edge interiors determine a finite subdivision \(G'\). Forward each expanded path segment by segment through all subdivision vertices it encounters, and combine the operations of abstract nodes at the same location. This gives a scheme of the same cost on \(G'\). By Lemma~\ref{lemma:subdivision} we obtain a scheme on \(G\) of no greater cost. Taking infima proves~\eqref{eq:graph_equality}.
\end{proof}

Together with the two maps of Theorem~\ref{theorem:two_sided_maps}, the four lemmas above yield Theorem~\ref{theorem:surface_cost}.

\begin{proof}[Proof of Theorem~\ref{theorem:surface_cost}]
Take the smooth closed surface \(M\subset\mathbb{R}^3\) of genus \(m-n+1\) and the maps \(\iota:|G|\to M\) and \(\pi:M\to|G|\) from Theorem~\ref{theorem:two_sided_maps}. Set \(f=\iota|_V\). Since \(\pi(\iota(v))=v\) for all \(v\in V\), the map \(f\) is injective on \(V\), the map \(\pi\) is injective on \(f(V)\), and \(\pi(f\mathcal{D})=\mathcal{D}\). Applying Lemma~\ref{lemma:lipschitz_transport} to \(\pi\) with \(\operatorname{Lip}(\pi)\le1\) and to \(\iota\) with \(\operatorname{Lip}(\iota)\le1+\varepsilon\) gives

\begin{equation}\label{eq:collapse_transport_route}
  \Route_{|G|}(\mathcal{D}) \le \Route_M(f\mathcal{D}) \le (1+\varepsilon)\Route_{|G|}(\mathcal{D}),
\end{equation}

\begin{equation}\label{eq:collapse_transport_code}
  \Code_{|G|}(\mathcal{D}) \le \Code_M(f\mathcal{D}) \le (1+\varepsilon)\Code_{|G|}(\mathcal{D}).
\end{equation}

Replacing \(|G|\) by \(G\) in these bounds using Lemma~\ref{lemma:metric_graph} gives~\eqref{eq:route_surface} and~\eqref{eq:code_surface}. If \(\Code_G(\mathcal{D})>0\), then~\eqref{eq:code_surface} implies \(\Code_M(f\mathcal{D})\ge\Code_G(\mathcal{D})>0\). Dividing the corresponding bounds in~\eqref{eq:route_surface} and~\eqref{eq:code_surface} yields~\eqref{eq:advantage_transfer}.
\end{proof}

For example, the XOR-swap network in~\cite[Appendix~C]{Dysfunctor2026} has $32$ vertices and $84$ unit-length edges. Its optimal fractional routing cost is $126$, and a feasible coding scheme has cost $125$. Taking $0<\varepsilon<1/125$ in Theorem~\ref{theorem:surface_cost} gives a smooth closed surface of genus $84-32+1=53$ on which optimal network coding cost is lower than optimal fractional routing cost.

\begin{proof}[Proof of Theorem~\ref{theorem:unbounded_advantage}]
Fix $A>1$. Counterexamples on undirected multiple-unicast networks~\cite{ZhangLiLi2026,BravermanHe2026} achieve a higher common rate with network coding than with fractional routing. The construction in~\cite{BravermanGargSchvartzman2017} then gives a family of graphs on which the ratio of the maximum common rates is arbitrarily large. By throughput–cost duality~\cite[Theorem~4.1]{XiahouLiWuHuang2014}, there exist a finite connected positively weighted undirected graph $G$ and a demand family $\mathcal{D}$ with $\Adv_G(\mathcal{D})>2A$. Taking $\varepsilon=1$ in Theorem~\ref{theorem:surface_cost} gives a surface $M$ and corresponding demands $f\mathcal{D}$ satisfying

\[
\Adv_M(f\mathcal{D})\ge \tfrac12\Adv_G(\mathcal{D})>A.\qedhere
\]
\end{proof}

\subsection{Coding Advantage at Fixed Genus}

We prove the \(O(\log(g+2))\) bound on coding advantage for graphs of orientable genus at most \(g\) and smooth closed surfaces of genus at most \(g\) in two steps. First we embed the shortest-path metric of a bounded-genus graph into \(L_1\) and decompose the resulting metric into a nonnegative combination of cut metrics. The information-theoretic cut bound then gives the cost inequality on graphs. A small perturbation of the links of any finite surface code, followed by insertion of forwarding nodes at crossings, reduces the surface case to a finite graph embedded without crossings in that surface.

\begin{proof}[Proof of Theorem~\ref{theorem:genus_bound}]
If \(\mathcal{D}\) is empty, both costs vanish and~\eqref{eq:genus_advantage_bound} holds. Assume henceforth that \(\mathcal{D}\) is nonempty.

First consider a finite connected undirected simple graph \(H=(V_H,E_H,\ell)\) of orientable genus at most \(g\), with positive edge lengths \(\ell_e\) and shortest-path metric \(d_H\). The constant-distortion embedding theorem for planar graph metrics~\cite[Theorem~1.1]{OpenAIPlanar2026} and the stochastic planarization corollary~\cite[Corollary~I.3]{Sidiropoulos2010} imply that \((V_H,d_H)\) embeds into \(L_1\) with distortion at most \(D=C\log(g+2)\) for an absolute constant \(C>0\). For \(g=0\), the planar bound applies directly. After rescaling, the induced \(L_1\) metric \(\rho\) on \(V_H\) satisfies \(d_H(x,y)\le\rho(x,y)\le D\,d_H(x,y)\) for all \(x,y\in V_H\). Since every finite \(L_1\) metric is a nonnegative linear combination of cut metrics~\cite{DezaLaurent1997}, there are weights \(w_S\ge0\) for \(S\subset V_H\) such that \(\rho=\sum_{S\subset V_H}w_S\delta_S\), where \(\delta_S(x,y)=1\) if exactly one of \(x,y\) belongs to \(S\) and \(\delta_S(x,y)=0\) otherwise.

Fix any feasible network code for \(\mathcal{D}\) on \(H\), and let \(b_e\ge0\) be the total bandwidth used on edge \(e\in E_H\) in both directions. For each vertex subset \(S\subset V_H\), all independent messages originating in \(S\) and demanded in \(V_H\setminus S\) must be recoverable from the transmissions entering \(V_H\setminus S\) together with the messages originating there. Hence the total bandwidth directed from \(S\) into \(V_H\setminus S\) is at least the sum of the rates of those sessions. Applying the same argument in the reverse direction and adding gives the two-way cut bound

\begin{equation}\label{eq:cut_bandwidth}
  \sum_{i=1}^k r_i\delta_S(s_i,t_i) \le \sum_{e=\{x,y\}\in E_H} b_e\delta_S(x,y).
\end{equation}

Multiply both sides of~\eqref{eq:cut_bandwidth} by \(w_S\ge0\), sum over all \(S\subset V_H\), and use \(d_H(x,y)\le\ell_e\) for each edge \(e=\{x,y\}\) together with Lemma~\ref{lemma:routing_cost} to get

\begin{equation}\label{eq:genus_graph_cost}
  \begin{aligned}
  \Route_H(\mathcal{D}) &= \sum_{i=1}^k r_i d_H(s_i,t_i) \le \sum_{i=1}^k r_i\rho(s_i,t_i) \\
  &\le \sum_{e=\{x,y\}\in E_H} b_e\rho(x,y) \le D\sum_{e\in E_H} b_e\ell_e.
\end{aligned}
\end{equation}

Taking the infimum over all feasible coding schemes on \(H\) gives \(\Route_H(\mathcal{D})\le D\,\Code_H(\mathcal{D})\).

Now let \(M\) be a smooth connected closed orientable Riemannian surface of genus at most \(g\). Fix a finite coding scheme on \(M\) meeting \(\mathcal{D}\) at cost \(V_0\), and fix \(\delta>0\). Merge abstract nodes at the same point of \(M\) and remove zero-bandwidth links and links with coincident endpoints. Write \(J\) for the set of remaining arcs, with bandwidths \(b_j>0\), \(j\in J\). Choose \(\tau>0\) such that \(\tau\sum_{j\in J}b_j<\delta\). Draw each arc \(j\in J\) as a minimizing geodesic between its endpoints on \(M\), and perturb the curves smoothly on \(M\). First separate incident arcs near their common endpoints, then perturb the interiors so that each resulting curve \(\alpha_j\) has length less than the intrinsic distance between its endpoints plus \(\tau\). Its interior avoids all existing node positions, and away from shared endpoints the curves meet only in finitely many transverse double crossings.

At each interior crossing insert a relay that forwards each link's transmissions along its own curve, splitting the family \(\{\alpha_j\}_{j\in J}\) into subarcs with disjoint interiors. To remove any parallel edges or loops, place one forwarding node inside each subarc with distinct endpoints and two distinct forwarding nodes inside each subarc whose endpoints coincide. Subdivision preserves total cost. Taking all nodes, including isolated terminals, as vertices and the subdivided arcs with their arclengths as undirected edges gives a finite undirected simple graph \(\widehat{G}\) embedded without crossings in \(M\). The original code runs along these subdivided arcs at total cost less than \(\sum_{j\in J}b_j(d_M(p(j^-),p(j^+))+\tau)<V_0+\delta\).

Write \(H_1,\dots,H_q\) for the connected components of \(\widehat{G}\). Each \(H_t\) embeds without crossings in \(M\), so its orientable genus is at most \(g\). Since each session has positive target rate \(r_i\), feasibility requires its source \(s_i\) and sink \(t_i\) to be connected by a transmission path. They therefore lie in the same component, partitioning the session indices into sets \(I_t\), one for each component \(H_t\), with corresponding demand subfamilies \(\mathcal{D}_t\). Each path in \(H_t\) is a piecewise smooth curve of the same length in \(M\), so \(d_M(s_i,t_i)\le d_{H_t}(s_i,t_i)\) for every session in \(\mathcal{D}_t\). Applying Lemma~\ref{lemma:routing_cost} and the graph bound to each component and summing gives

\begin{equation}\label{eq:genus_surface_cost}
  \begin{aligned}
  \Route_M(\mathcal{D}) &= \sum_{t=1}^q \sum_{i\in I_t} r_i d_M(s_i,t_i) \le \sum_{t=1}^q \Route_{H_t}(\mathcal{D}_t) \\
  &\le D\sum_{t=1}^q \Code_{H_t}(\mathcal{D}_t) \le D(V_0+\delta).
\end{aligned}
\end{equation}

Let \(\delta\to0\) and then take the infimum over all finite coding schemes on \(M\) to get \(\Route_M(\mathcal{D})\le D\,\Code_M(\mathcal{D})\).
\end{proof}

%% file: sections/conclusion.tex
\section{Conclusion}\label{sec:conclusion}

We proved that for every \(\varepsilon>0\), the shortest-path metric on the vertices of any finite connected undirected simple graph with \(n\) vertices and \(m\) positively weighted edges admits a noncontractive embedding with distortion at most \(1+\varepsilon\) into a smooth connected closed orientable surface in \(\mathbb{R}^3\) of genus \(m-n+1\). This almost-isometric embedding extends to every finite metric space. For the manifold hypothesis in data analysis, this means that when the topology and metric of the surface may vary with the sample, the intrinsic distances on a smooth two-dimensional surface can approximate any finite sample metric with arbitrarily small multiplicative error. In the space information flow model, for any finite connected positively weighted undirected graph and any $\varepsilon>0$, we construct a smooth closed surface such that for every multiple-unicast demand family on the graph, the optimal fractional routing and network coding costs on the surface each lie between the corresponding graph cost and $1+\varepsilon$ times that cost. Applying these cost bounds to known undirected graph instances with arbitrarily large coding advantage yields a family of smooth closed surfaces on which the ratio of optimal fractional routing cost to optimal network coding cost can be arbitrarily large. When the orientable genus of the graph or the genus of the surface is at most \(g\), coding advantage is at most \(C\log(g+2)\) for an absolute constant \(C\).

The maps $(\iota,\pi)$ connect the entire metric graph $|G|$ to the smooth closed surface $M$. Mapping a graph path to the surface by $\iota$ increases its length by at most a factor of $1+\varepsilon$, while $\pi$ takes arbitrary auxiliary points on the surface and the piecewise smooth curves joining them back to the graph without increasing curve length. Thus, for problems whose costs are nonnegative weighted sums of distances or curve lengths, optimal costs on the graph and the surface agree within a factor of $1+\varepsilon$ whenever feasible configurations transfer in both directions with their weights preserved. The transfer of costs and coding advantage in space information flow illustrates this mechanism, and the same reduction may apply to other problems linking discrete graphs and smooth manifolds.

Three questions remain about manifold embeddings and space information flow. First, statistical testing of the manifold hypothesis and geometric reconstruction typically impose regularity bounds on quantities such as curvature, reach, injectivity radius, or area. Under prescribed bounds of this kind, which finite metric spaces still admit almost-isometric embeddings into smooth surfaces? Second, can the \(O(\log(g+2))\) upper bound on coding advantage at fixed genus \(g\) be improved? Third, multiple-unicast coding advantage equals \(1\) in \(\ell_p^n\) for \(1\le p\le 2\), including Euclidean space, and exceeds \(1\) on some smooth closed surfaces. In which other metric spaces or Riemannian manifolds is multiple-unicast coding advantage always \(1\), and which geometric properties determine this distinction?

%% file: sections/acknowledgment.tex
\begin{samepage}
\section*{Acknowledgment}

The research problems, core ideas, and overall proof framework of this paper were conceived and designed by the authors. The authors used OpenAI ChatGPT to help draft mathematical text and refine mathematical expressions, and used AI tools including Google Gemini to edit and polish the language of the abstract and Sections~\ref{sec:introduction}--\ref{sec:conclusion}. The authors also used OpenAI Codex to write the TikZ code for Figs.~\ref{fig:three_steps} and~\ref{fig:tube_maps}. The authors verified all mathematical proofs, technical details, and final text, and take full responsibility for the content of this paper.\par
\end{samepage}

%% file: references.tex
\bibliographystyle{IEEEtran}
\bibliography{references}